\documentstyle[preprint2]{aastex}

\begin{document}

\noindent {\it Conference Highlights}

\begin{center}

\title{\large \bf Angular Momentum Evolution of Young
Stars: Toward a Synthesis of Observations, Theory, and
Modeling\footnote{The conference was held in Albuquerque,
New Mexico, 2002 June 5, as a Topical Session at the 200th
Meeting of the American Astronomical Society.}}

\end{center}

\medskip

The aim of this AAS Topical
Session was to update the community on the current state of
knowledge about the angular momentum evolution of
young stars. For newcomers to the subject, the session was
intended to provide an introduction and general overview,
and to highlight emerging issues.  For experienced workers
in this field, the session provided an
opportunity for synthesizing recent developments in
observations, theory, and modeling of rotation of
young stars, and for identifying promising new research
directions.


\section{Overview of Angular Momentum Evolution of Young 
Stars---{\it Donald Terndrup (Ohio State University)}}

During the last decade we have witnessed a tremendous
growth in the amount and quality of data on stellar
rotation, which now includes samples in ages from 
T Tauri stars to clusters as old as the Hyades.  Along
with this has been an increase in the sophistication
of theoretical models of the spindown of stars before and on
the main sequence.  Many of us in the field now believe
we have a good handle on the three principal ingredients
of these models:

$\bullet$ The distribution of initial rotation rates in the
pre-main sequence, coupled with the (possible) influence
of accretion disks.

$\bullet$ Angular momentum loss from a magnetized solar-like
wind, which operates on time scales of tens to hundreds
of Myr.

$\bullet$ Prescriptions for the angular momentum distribution
in convective regions of the star and angular
momentum transport in radiative regions.

The initial conditions for rotation are set during the star
formation process.  T Tauri stars are logical starting
points, as they represent the earliest easily-observable
phases of evolution at which stars have reached their
final mass.  These stars are at the beginning of the
hydrostatic phase of evolution prior to the ignition of
nuclear reactions, and typically have modest accretion
disks of order 0.01 M$_\odot$.  There is extensive
data on rotation rates of T Tauri stars,
beginning with the work of Vogel \& Kuhi (1981), which shows
that there is already a significant range of rotation rates
among protostars with ages of 1 Myr or less.  Recent
surveys in the Orion Nebula Cluster (ONC), which has an age
of 1--3 Myr, have provided hundreds of rotation
periods as will be reviewed below.

The observation that the youngest stars rotate at a small
fraction (10--20\%) of the breakup velocity indicates that
they must have had an efficient method for shedding angular
momentum.  The most common approach to modeling this is
to invoke ``disk-locking,'' as introduced by K\"onigl
(1991).  In the presence of an accretion disk, a star could
maintain a nearly constant surface rotation rate as it
contracts (Shu et al.\ 1994); this corotation between the
star and the accretion disk would govern the stellar
rotation rate.	Theoretical models of angular momentum
evolution have therefore included the accretion disk
lifetime as a parameter describing the initial conditions
(Sills, Pinsonneault, \& Terndrup 2000 and references therein).

Later on, when stars have arrived on the main sequence,
the stars experience continuing  angular momentum loss,
envisioned as resulting from a magnetized stellar wind
(Weber \& Davis 1967).  In all young clusters, most stars
are rotating slowly, with $v \sin i < 20$ km s$^{-1}$, but a
minority are rotating must faster.
The loss mechanism efficiently removes stellar angular
momentum: in clusters such as Alpha Persei (age 50 -- 80
Myr), the fastest rotators have equatorial velocities near
200 km s$^{-1}$, but these have fallen to below 10
km s$^{-1}$ by the
age of the Hyades (Radick et al.\ 1987).  The existence of
rapid rotation at ages of $\sim 100$ Myr suggests that the loss
rate is saturated or suppressed at high rotation rates.
Observations of stellar activity (e.g., Krishnamurthi et
al.\ 1998) and a variety of theoretical arguments lead to
the idea of a saturation threshold (Kawaler 1988).  In this
picture the loss rate scales as $\omega^3$ 
for slow rotators (Skumanich 1972), 
where $\omega$ is the angular velocity,
but scales as $\omega^2_{\rm crit}\omega$ 
for $\omega > \omega_{\rm crit}$.  The saturation threshold
depends on stellar mass, in the sense that it has a lower
value for stars of lower mass (Krishnamurthi et al.\ 1997).

The internal distribution of angular momentum within a star
must also be understood in a complete theory of angular
momentum evolution.  Low-mass stars are initially fully
convective.  Their surface rotation rates can be used to
infer an initial angular momentum if the angular momentum
distribution in convective regions is specified, since
the convective overturn timescale is much shorter than
the evolutionary timescale.  Once on the main sequence,
however, stars with masses above 0.35 solar masses develop
radiative cores.  Stars with higher masses have much
thinner convective zones:  at 1 solar mass, for example,
only 10\% of the moment of inertia is in the convective
envelope.  When a radiative core develops, the internal
transport of angular momentum in the radiative regions must
be accounted for, since the timescales for evolution and
for angular momentum transport are comparable.	All this
means that the models have to begin with assumptions
about the (radial) angular momentum profile; obvious
limiting cases are to assume initial solid body rotation
or constant specific angular momentum.	The models also
need to have an adjustable efficiency of internal angular
momentum transport. 

\section{Rotation of Solar-Mass Stars at Different
Stages of Evolution: Observations}

\subsection{Rotation of Protostars---{\it Thomas
Greene (NASA Ames) \& Charles Lada (CfA)}} 

It is important to determine the
angular momenta of embedded protostars
to understand better the star-formation accretion process
and to establish the angular momentum ``zero point'' from
which stars evolve.  Unfortunately, the angular momenta of
protostars are largely unknown. Very high extinctions ($A_V
\geq 40$ mag) make them impossible to observe in visible
light with high-resolution spectrographs. Further, these
objects are not very bright in the near-IR ($K \sim 10$),
and they have very large continuum veilings ($r \sim 3$)
at these wavelengths, greatly diluting the strength of
their photospheric absorption lines and star spots.

Despite these difficulties, we have recently obtained
high-resolution absorption spectra of some protostars
using sensitive IR spectrographs on large telescopes. These
data show that protostars with flat-spectrum and Class~I
spectral energy distributions (SEDs) have spectral types,
surface gravities, and radii similar to more evolved
T~Tauri stars (TTS).  However, the embedded protostars
observed to date have considerably different $v \sin i$
and angular momenta. These objects appear to be rotating
significantly faster than most TTS; typical protostars have
$v \sin i \sim 40$ km s$^{-1}$ (Greene \& Lada 1997, 2002).
Recent ASCA and Chandra X-ray observations also suggest
that these protostars have short rotation periods.

\begin{figure}[ht] 
\plotone{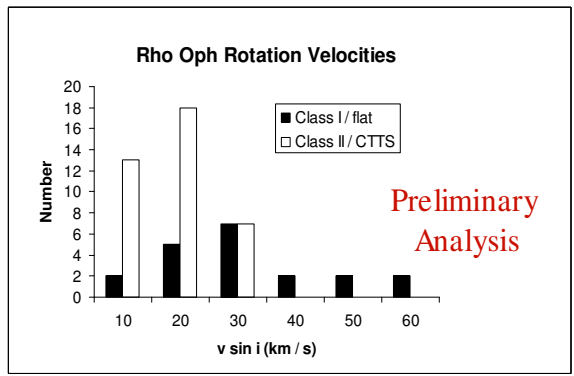} 
\caption{Rotation velocities for T Tau stars in the $\rho$ Oph
region.  Class I objects and those with flat spectral energy
distributions have faster rotation than class II 
and classical T Tau stars.} 
\end{figure}

Our Keck NIRSPEC observations of YLW~15 have provided the
first measurement of the angular momentum and detailed
astrophysics of a Class~I protostar (Greene \& Lada
2002). With a projected rotation velocity $v \sin i = 50$ km
s$^{-1}$, its co-rotation radius is about 2 $R_\star$. This
is about equal to the magnetic coupling radius predicted
by the Shu et al.\ (1994) X-wind/magnetospheric-accretion
model for the observed mass accretion rate of $\dot{M}
= 1.6 \times 10^{-6} M_\odot$ yr$^{-1}$.  Like other
Class~I protostars, YLW~15 will evolve into a TTS in
$\sim 10^5$ yr. Since TTS in the same $\rho$ Oph cloud
rotate only about one-third as fast (e.g., Greene \&
Lada 1997), YLW~15 (and other similar protostars) must
shed approximately two-thirds of its angular momentum
in about this amount of time. Within the framework of
magnetic star-disk interaction, this is possible provided
that its mass accretion rate drops quickly to $10^{-7}
M_\odot$ yr$^{-1}$.

These observations are in their infancy, but they
may prove essential for understanding initial stellar
angular momentum and star-disk interactions during the
protostellar evolutionary phase. Soon we expect to extract
statistically significant results on these issues from
our recently-completed Keck NIRSPEC observations of a few
dozen embedded protostars in nearby star-forming clouds.

\subsection{Rotation of Pre--Main-Sequence Stars---{\it
William Herbst (Wesleyan University)}}

Low-mass stars often have
large, stable, non-axisymmetric, cool spots on their
surfaces which makes it possible to measure their rotation
periods by photometric monitoring. There is nothing we
can determine with as much accuracy for PMS stars as their
rotation rates.

Angular velocities accurate to about 1\%
are currently available for $\sim 1500$ PMS stars (see
Fig.\ 2). Approximately 40\% of the stars with $12.5 <
I < 15.5$ in the Orion Nebula Cluster (ONC), the best
studied example, now have rotation periods measured
for them. 
There is no reason to believe that this
sample is significantly biased with respect to rotation
properties (Rhode et al.\ 2001; Herbst et al. 2002) or
other characteristics, so it provides an excellent means
of studying rotation among 
very young stars.   We focus our discussion on
angular velocity instead of angular momentum, because
the latter requires knowledge of the radii of stars,
which are particularly uncertain at PMS ages.

\begin{figure}[ht] \plotone{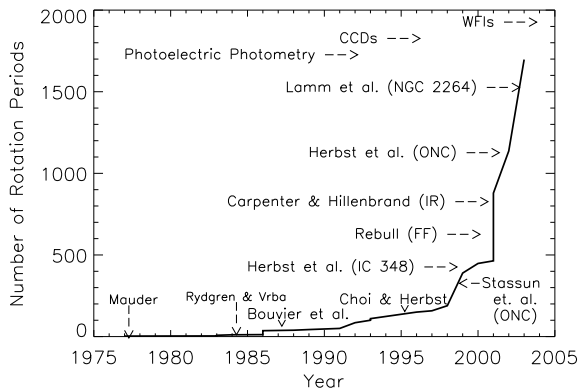} \caption{The
spectacular growth in numbers of PMS stars with measured
rotation periods over the last few years is primarily due
to the recent availability of wide field CCD imagers on
moderate sized telescopes to which sufficient access can
be gained to permit monitoring studies.} \end{figure}

The first surprising fact about PMS rotation rates is
the breadth of the distribution---it extends over at
least a factor of 30 from $\sim 2/3$ day to $\sim 20$
days. The second surprising fact is that it is strongly
mass dependent. Stars in the range of $\sim 1/4$ to 1
$M_\odot$ have a bimodal distribution (Fig.\ 3) with peaks near 2
d and 8 d, as first shown by Attridge \& Herbst (1992)
and Choi \& Herbst (1996), while lower mass stars have a
unimodal distribution and generally spin faster (Herbst et
al.\ 2001). This behavior is not limited to the ONC, but
is seen also in NGC~2264 (Lamm et al.\ 2002).  As NGC~2264
is a somewhat older cluster, the peaks in the higher mass
distribution are at 1 d and 4 d, instead of 2 d and 8 d as
in the ONC.  This is evidence for spin-up of a significant
percentage of the PMS stars in NGC~2264. Assuming that it
is three times the age of the ONC (i.e.\ 3 My as opposed
to 1 My), the amount of spin-up (i.e.\ a factor of 2)
is consistent with expectation based on PMS contraction
models and conservation of angular momentum.

\begin{figure}[ht] \plotone{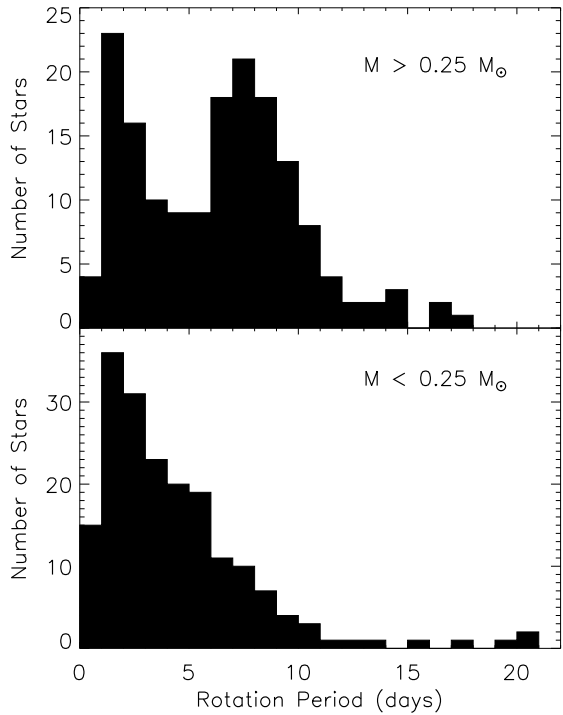} \caption{The period
distribution of ONC stars for two mass ranges. (Taken from Herbst
et al.\ 2002.) } \end{figure}

It is interesting
to speculate about how these (mass dependent) distributions
could have arisen during the early PMS or proto-star
stages.  The leading hypothesis is disk-locking. One
possibly relevant piece of information is that, in the ONC,
we find a distinct correlation between rotation rate and
strength of near-IR excess (indicative of the presence of
an inner circumstellar disk). In particular, the $I-K$
excess emission is $0.55 \pm 0.05$ for slowly rotating
stars (i.e.\ those with angular velocities less than 1,
corresponding to rotation periods longer than 6.3 days),
while it is $0.17 \pm 0.05$ for rapid rotators (with
$\omega > 2$ corresponding to $P < 3.1$ days). The first
order interpretation of this is that disks tend to
brake stars, as the disk-locking theory predicts. However,
under some scenarios one expects a correlation
in the opposite sense, as Stassun et al.\ (2001) have
discussed. A deeper understanding of the meaning of the
near-IR correlation and its relation to disk-locking
theory will require further study, particularly aimed at
establishing the presence and nature of disks around PMS
stars with known rotation rates.

\subsection{Rotation of solar-type stars at zero age and
beyond---{\it Sydney A.\ Barnes (University of
Wisconsin, Madison)}}

The most basic and most solidly known result in stellar
rotation is that of the solar rotation rate, on the
surface, and more recently, in the interior.  We know
that upper main sequence (F and earlier) stars spin fast,
and that lower main sequence (G, K \& M) stars spin
slowly. This is related respectively to the absence and
presence of surface convection zones and winds on these
stars (Schatzman 1962).  

For lower main sequence stars, the spindown is described by
the Skumanich (1972) relationship: $v \propto t^{-1/2}$,
where $v$ and $t$ are the stellar rotation velocity and
the age of the star respectively. This suggests that on
the main sequence, the rate of angular momentum ($J$)
loss is related to the angular velocity ($\omega$)
by ${dJ}/{dt} \propto \omega^3$. Apparently, this
relationship breaks down among very fast rotators,
a phenomenon referred to as magnetic ``saturation".

But there are complications when one deals with young
stars.  It appears that
in addition to contracting on the PMS, a star can also
interact with its circumstellar disk, and this can have
profound effects on the rotational evolution. Once on the
main sequence, winds dominate the rotational evolution,
except that internal transport of angular momentum can
also play an important role.

Thus, the present framework for interpreting the
observations involves: pre-main sequence evolution,
magnetic saturation, disk interaction, winds, and internal
transport.  The complexity of these phenomena requires
observational guidance to decode.  On the PMS, 
a wealth of data has been provided by Herbst and
collaborators, Stassun et al.\ (1999) and Rebull (2001).
On the main sequence, 
two $v \sin i$ datasets have
emerged recently, for the open clusters M\,34 (Soderblom
et al.\ 2001) and NGC\,2516 (Terndrup et al.\ 2002). With
ages of approximately 250\,Myr and 150\,Myr respectively,
these observations are helping to fill in the gap in the
observations between the young clusters ($\le$\,100\,Myr)
and the Hyades (600\,Myr).

Today, significant numbers of rotation
periods among main sequence stars are available\footnote{We
reference here only the largest datasets.} in IC\,2391
(Patten \& Simon 1996), IC\,2602 (Barnes et al.\ 1999),
IC\,4665 (Allain et al.\ 1996), Alpha\,Per (Prosser \&
Grankin 1997; Prosser et al.\ 1993), the Pleiades (van
Leeuwen et al.\ 1987; Krishnamurthi et al.\ 1998), and
Hyades/Coma (Radick et al.\ 1987; Radick et al.\ 1990),
spanning an age range from 30\,Myr to 600\,Myr.  In the
immediate future, we can expect the release of datasets
in NGC\,1039 (M\,34), NGC\,2516, and NGC\,3532 (Barnes and
collaborators), two of which were displayed in preliminary
form at the June 2002 AAS meeting.  The author is aware
of on-going studies in M\,34 and M\,35 (S.\ Meibom,
personal communication).  These and other observations
will ensure that it is possible to build a time sequence
for a range of stellar masses, thus completely removing
the ambiguities inherent in $v \sin i$ data.

It is already possible to identify trends in
the complete dataset. These were pointed out during the
meeting, and will be published in the immediate future.
The color-magnitude diagrams of the
newly studied clusters, often generated in conjunction
with membership studies of one sort or another, appear
to be very promising, and enable the identification even
of very low mass cluster members. Trends in the amplitudes of rotational
variability are slowly emerging, and we are beginning to
decode its dependence on stellar mass, age, and rotation
rate.

Lest implications on the PMS and during the early main
sequence evolution distract us, it is worth noting that
the Skumanich (1972) relationship is still an excellent
description of main sequence stellar spindown, once the
stars have been binned by stellar mass. This is explained
at length in Barnes (2001), which considered whether and
how the rotation rates of the extremely well-studied
sample of Mt.\,Wilson stars can be connected to those
in open clusters, and whether the rotation rates of the
host stars of extra-solar planets could be distinguished
from the others.  The result is that, with a few minor
exceptions that can be explained, a Skumanich-type spindown
is a perfectly adequate description of the behavior of
both the Mt.\,Wilson stars and of the planet host stars.
The planet host stars are not particularly slow rotators,
in contrast to expectations from disk-locking scenarios
of stellar rotation.

\section{Rotational Evolution of Solar-Mass Stars: Theory}

\subsection{Internal Angular Momentum Transport
Processes---{\it Marc Pinsonneault (Ohio State
University)}}


Hydrodynamic mechanisms (meridional circulation, shear, and
GSF instabilities) can cause mixing and they are effective
in rapidly rotating stars.  However, these mechanisms occur
over a long timescale in slowly rotating stars.  Turbulence
in convective regions can excite waves where the restoring
force is variations in the potential ($g$-modes); such waves
can effectively transport angular momentum throughout
the radiative cores of stars.  Significant mixing would
only occur near the edges of convection zones.  However,
the impact of this mechanism upon stellar models has been
controversial and the overall impact on the behavior of
stars is not clear.  Finally, if there exists a magnetic
field in the radiative core of a star it will tend to
enforce corotation along field lines without inducing
significant mixing.  The impact upon stellar properties,
however, depends significantly upon the overall morphology
of the field.

We have two important constraints on the timescale for
angular momentum transport in stars.  Helioseismic data
provide a measurement of the internal solar rotation
as a function of radius and latitude.  There is a shear
layer located beneath the convection zone with a
characteristic width of less than 0.05 solar radii, and the
radiative core below the shear layer appears to rotate as a
solid body.  This indicates that angular momentum transport
in the solar interior was highly effective, but it does not
directly provide an indication of the relevant timescale.
The second constraint is the spindown of slowly rotating
stars.  Because only a fraction of the total moment
of inertia is contained in the outer convection zone,
there will be a difference in the time evolution of the
surface rotation rates between models where the timescale
for internal angular momentum transport is short (e.g.\ the
whole star must be spun down by the wind) and models where
the timescale for internal angular momentum transport is
long (e.g.\ initially only the convection zone is spun down,
and the core is coupled at a later epoch.)

Models which include only angular momentum transport
from hydrodynamic mechanisms predict that as stars lose
angular momentum from a magnetized wind there will be an
initial phase where the envelope spins down with respect
to the core.  At later ages the models will reach a rough
equilibrium where the core is rotating significantly faster
than the surface.  The observed spindown of the slow
rotator population in young open clusters is consistent
with the early core-envelope decoupling predicted by
such models, but the flat solar rotation profile is
inconsistent with the strong internal angular velocity
gradients predicted by such models. This combination
of evidence suggests that there is an additional angular
momentum transport mechanism that operates over timescales
longer than 100 Myr but shorter than the lifetime of the
Sun.  Possible methods for distinguishing between waves
and magnetic fields were also briefly discussed.

\subsection{Magnetic Star-Disk Coupling Mech-anisms---{\it
Frank Shu (Tsing Hua Univ.)}}\footnote{Written by K.\ Stassun from notes 
provided by F.\ Shu}

The idea of 
magnetic ``disk-locking" in some sense originates with the
theory developed by Ghosh \& Lamb (1979) in the context of
neutron stars. In their model, a
stellar dipole field, $B$, everywhere threads a circumstellar
disk, which is truncated at an inner radius, $R_t$, where 
magnetic and ram pressures balance. In this picture, if $R_t$ 
is within the co-rotation radius, $R_c$ (the radius in 
the disk that rotates with Keplerian angular velocity equal
to the stellar angular velocity), then gas interior to $R_c$
spins up the star while gas exterior to $R_c$ spins it down.

But the Ghosh \& Lamb picture presents some problems in the
context of young, low-mass stars. By construction, 
the magnetic field is strong enough marginally to compete
with centrifugal forces in the funnel flow. Thus, since the disk
must be much more dense than the funnel flow, the field
threading the disk cannot resist wrapping except at the single
radius $R_c$. Continuous wrapping of the field would lead
to magnetic reconnection, but the observed X-ray activity of
classical T Tauri stars (those with disks) is similar to that
of weak-lined T Tauri stars. 

Our own X-wind model avoids these difficulties
by bringing the dipole field in the disk together at the
X-point, $R_x \approx R_c$. All of the unperturbed dipole
exterior to $R_x$ is brought in by disk accretion to $R_x$,
and half of the interior field is brought to $R_x$. This pinch
towards $R_x$ is caused by back-reaction to the funnel flow
and X-wind. This pinch distorts the dipole configuration near
$R_x \approx R_c$ so that differential rotation of the disk no
longer wraps the field, even though the disk is much denser
than the gas in the flow/X-wind, making a steady state 
configuration possible.

The ``dipole" assumption has problems when confronted by
observation, however. For example, the results of 
polarization studies are inconsistent with well-organized
fields at the stellar surface, and the hot spots that have
been observed on classical T Tauri stars have much smaller
areas than implied by the dipole assumption. Furthermore,
completely convective stars are not expected to have dipole
field geometries near the surface. 

Fortunately, X-wind theory is not dependent on the 
(unperturbed) magnetic field being dipolar. The critical
invariant in the theory is the assumption of ``trapped
magnetic flux" at the X-point, $R_x$. Relaxing the dipole
assumption, we find that ``disk-locking" in the X-wind model
is only very slightly modified in the ``multipole" case, 
but the model now allows for the smaller hot spots and 
polarization levels that are observed. 

We make two critical assumptions:

\noindent (1) Disk locking: $\Omega_\star = \Omega_x$, and

\noindent (2) Flux trapping: 
$\Phi_t = \alpha (G M_\star \dot{M}_D / \Omega_\star)^{1/2}$,

\noindent
where $\Phi_t$ is the magnetic flux at the truncation radius,
and $\alpha$ is an order unity coefficient that depends on 
details of X-wind theory.

In the X-wind model, the trapped flux is 1/3 in the outflowing
X-wind, 1/3 in the ``dead zone" (static coronal gas), and
1/3 in the funnel flow. Since the 1/3 of the trapped flux
in the funnel flow is anchored in the stellar surface, we
have the prediction that:

\centerline{$B (f 4 \pi R_\star^2) = 1/3 \Phi_t$,} 

\noindent where $B$ 
is the field strength in the footpoint on the stellar
surface, and $f$ is the surface covering fraction of that 
footpoint. Assuming that $B$ does not vary much from star
to star, this leads to the testable prediction that:

\centerline{$f R_\star^2 \propto (M_\star \dot{M}_D P_{\rm rot})^{1/2}$,}

\noindent where $P_{\rm rot} = 2 \pi / \Omega_\star$.

Recently, Johns-Krull \& Gafford (2002) compared observations
of T Tauri stars in Taurus-Auriga to the ``disk-locking"
prediction of this generalized, ``multipole" X-wind model. 
Their results, shown in Fig.\ 4, seem to verify this
prediction.

\begin{figure}[ht]
\plotone{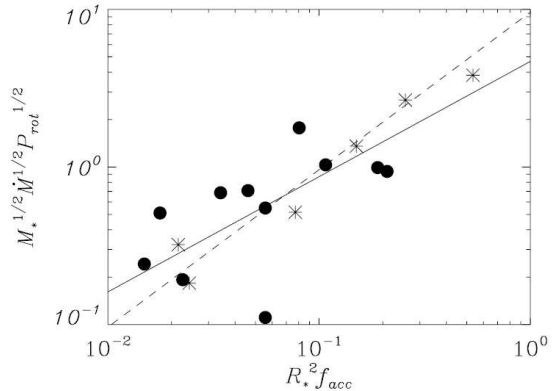}
\caption{The quantity $(M_\star \dot{M}_D P_{\rm rot})^{1/2}$
vs.\ $f R_\star^2$ for Tau-Aur stars with the requisite data. 
Asterisks represent binaries. The solid line is the best-fit
line to the data, while the dashed line represents the 
theoretically predicted relationship.
(Taken from Johns-Krull \& Gafford 2002).}
\end{figure}

We close by commenting on the eventual breakdown of disk
locking. At some point, as $\dot{M}_D$ decreases in time,
disk locking must break down, because the star has too much
inertia for the then-meager funnel flow to force the star to
rotate at $\Omega_\star = \Omega_x$. However, unlike the 
situation where the stellar field is a fixed dipole, disk
locking may have some flexibility built into it from the
ability of corotating multipole fields to accommodate even
fairly low disk accretion rates. Because of this complication,
and because we do not know how stellar dynamos work in the
presence of interactions with a surrounding accretion disk,
we have no first-principles, quantitative theory of disk
locking.

\section{Panel Discussion: Observations and Theory
vis-a-vis Modeling} 

The session included a panel discussion
which focused on a set of guiding questions posed by Keivan
Stassun (University of Wisconsin), who
moderated the discussion. The aim of the panel discussion
was for the observers, theorists, and modelers to begin
synthesizing work from one another's domains, to confront
areas of uncertainty, and to identify future research
directions.
The specific questions discussed by the panel were:

\smallskip
\noindent Disks: 
\begin{enumerate} \item How can we understand
the observed relationship between rotation and infrared
excesses [among PMS stars in Orion] in the context
of star-disk coupling?  

\noindent Observations in Orion show a 
{\it positive} correlation between rotation period and
magnitude of near-IR excess. This would seem to suggest
that the inner edges of the circumstellar disks in these
stars know something about the rotation rate of the star.
This is puzzling because near-IR emission probes the dust
in the disk, not the gas that is thought to magnetically
couple to the star. And, if any correlation might be
expected, it is in the opposite sense of what is seen.
The panel was not able to put closure
on this issue, and it remains unclear what this observational
result might imply for the physics of magnetic star-disk
interaction.

\item Does star-disk coupling
always act to {\it remove} angular momentum from a star?

\noindent Models of PMS angular momentum evolution typically
invoke ``disk-locking" as a way of extracting angular momentum
from the star. However, strictly speaking, accretion torques
can act positively on the star as well as negatively, 
depending on the balance between stellar magnetic field
strength and the accretion rate in the disk. Panelist Frank
Shu confirmed that in some cases stellar spin-up can result
from star-disk interaction.
\end{enumerate}

\smallskip
\noindent Winds: 
\begin{enumerate} \item What do we know (or wish we
knew) about solar-analog winds from young stars and their
role in early angular momentum evolution?  

\noindent Panelist Marc Pinsonneault suggested that more
quantitative observations of stellar magnetic field
strengths and of mass outflow rates would be extremely
helpful in more rigorously constraining models. Indeed,
there is some uncertainty about the basic relationship
between stellar rotation, dynamo-generated field strength,
and mass outflow rates. In addition, observations that
probe the internal rotation profiles of young stars are
needed, specifically to test the prediction that young
stars are strong differential rotators internally.
\end{enumerate}


With respect to future research directions, the following
were presented as key areas of observational focus:

$\bullet$ rotation measurements prior to 1 Myr 

$\bullet$ rotation measurements in the ``age gaps"
(primarily between 1 Myr and 100 Myr); connect rotation
period distributions at various ages 

$\bullet$ better
assessment of biases in measured rotation rates at various
ages 

$\bullet$ ascertain the role of stellar multiplicity
on the angular momentum evolution of single PMS stars

\section{New Frontiers at Other Masses: Inter-mediate-Mass
Stars and Brown Dwarfs}

\subsection{Intermediate Mass Stars: From the
Birthline to the Main Sequence---{\it Sidney C.\
Wolff (NOAO)}, {\it S.\ Strom (NOAO)},
{\it L. Hillenbrand (Caltech)}} 

We have measured
projected rotational velocities ($v\sin i$) for a sample
of  145 stars located in the Orion star-forming complex
with masses ($M$) between 0.41 and 15 M$_\odot$.  These
measurements establish, as nearly as direct observations
can, the initial values of specific angular momentum ($j
\equiv J/M$) for stars in this mass range.  The goal of
this work is to try to account for the initial values of
$j$ in terms of theories of star formation and angular
momentum regulation.  We then map the initial values of
$j$ onto the ZAMS and discuss the processes that affect
the evolution of angular momentum from the birthline to
the ZAMS.

Our data (see Fig.\ 5) show that there is a continuous power law
relationship between $j$ and $M$ for stars on convective
tracks with masses in the range $\sim 0.5$ (and probably
less) to 2--3 M$_\odot$ and that this power law merges
smoothly with the relationship for more massive stars,
which are already on the ZAMS.  This power law can be
compared with the predictions of models that posit that
stars are ``locked" to circumstellar accretion disks
until they are released at the birthline.  If we assume
that the accretion rate is $10^{-5}$ M$_\odot$ yr$^{-1}$
for 1 M$_\odot$ stars, and that mass accretion rates
are proportional to stellar mass, we can account for the
observed slope and zero point of the upper envelope of
the observed power law relationship between $j$ and $M$.

\begin{figure}[ht] 
\plotone{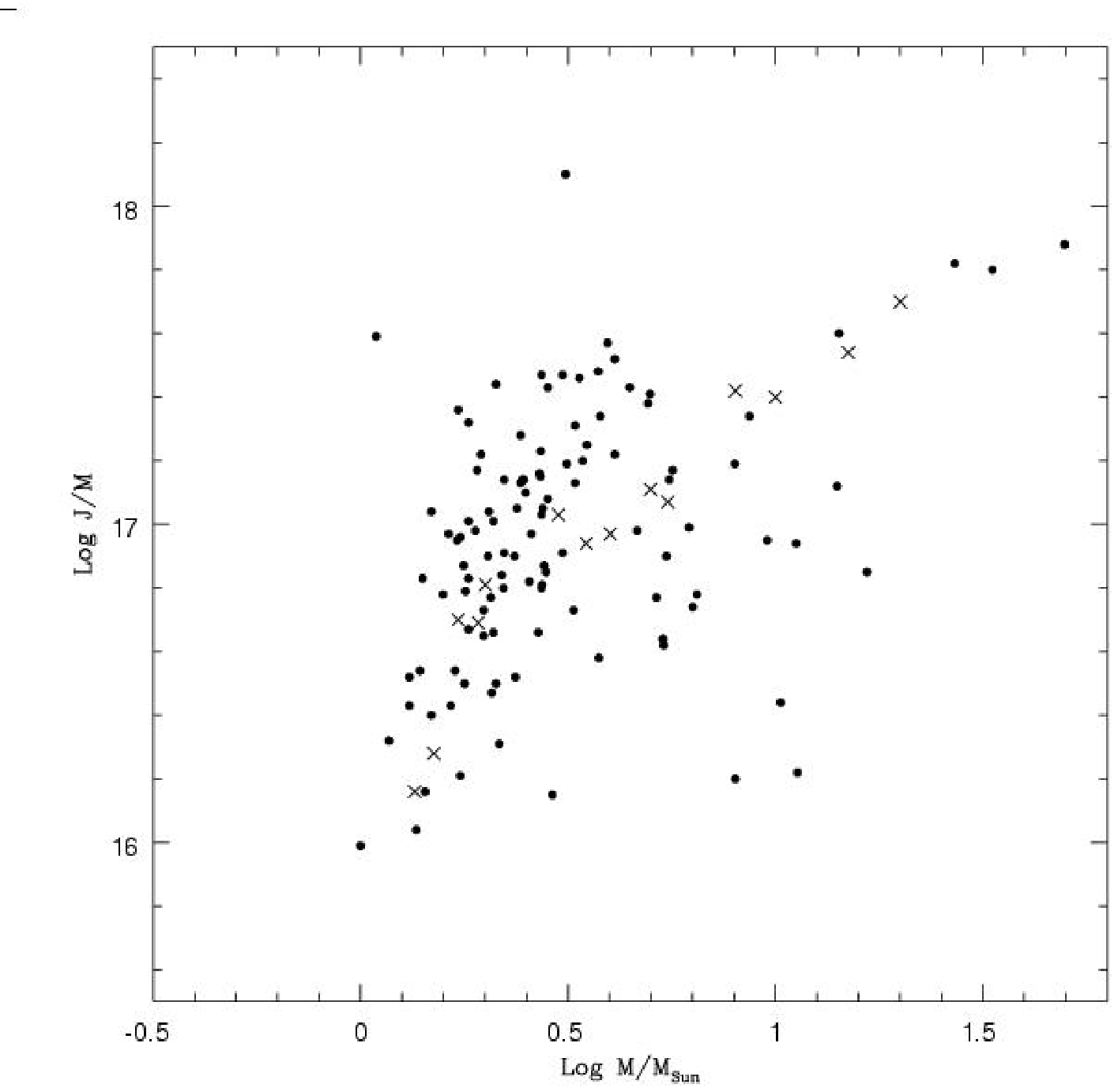} 
\caption{Observations of $J/M$ vs.\ $M$ for intermediate-mass
stars in the ONC.} 
\end{figure}

If we compare the initial values of $j$ with those of
stars on PMS radiative tracks or on the main sequence,
we find the primary difference to be a break in the power
law for stars with $M < 2$ M$_\odot$ that have completed
the convective phase of evolution.  This sharp decrease in
$j$ has been known for a long time for field stars, but the
decrease is seen already in the Orion stars, which are less
than 1 Myr old and are still on PMS radiative 
tracks.  We show that the break in the power law is a
consequence of core-envelope decoupling at the time of
the transition from the convective to radiative tracks.
The difference in rotation rates seen for stars on either
side of this transition are consistent with conservation
of angular momentum in shells and inconsistent with solid
body rotation.

We also argue that the systematic variations seen in
$\left< v \sin i \right>$ along the ZAMS for stars with
masses greater than 2--3 M$_\odot$, i.e.\ a maximum
in $\left< v\sin i \right>$ in the late B-type stars,
lower $\left< v\sin i \right>$ in the early B-type stars,
and a large fraction of B2--B4 stars rotating at less
than 100 km/s, can be accounted for by variations in the
predicted mass-radius relationship along the birthline,
primarily associated with deuterium shell burning, with
the amount of subsequent spin-up being determined by the
difference between the radius on the birthline and the
radius on the ZAMS.

\subsection{Rotation and Activity of Ultra-Cool
Dwarfs---{\it Adam Burgasser (UCLA)}} 

The study
of angular momentum evolution in ultra-cool dwarf stars
and brown dwarfs is just beginning, as these faint objects
have until quite recently eluded detection (for a recent
review on brown dwarf searches, see Basri 2000).  However,
early indications show a connection between rotation
and stellar activity quite opposite that seen in hotter
stars (Noyes et al.\ 1984).  

High-resolution $v\sin i$
measurements by Basri et al.\ (2000) and Reid et al.\
(2002) have shown that in the field, cool L dwarfs
(Kirkpatrick et al.\ 1999; Mart{\'{\i}}n et al.\ 1999)
are systematically faster rotators than M dwarfs, with no
L dwarfs observed to date rotating slower than $v\sin i = 10$
km/s.  At the same time, magnetic H$\alpha$ emission shows
a steep and steady decline in both frequency and strength
beyond spectral type M7 (Gizis et al.\ 2000; Burgasser et
al.\ 2002c), possibly due to the low ionization fractions in
the upper atmospheres of the cooler dwarfs (Mohanty et al.\
2002).  This implies that rapidly rotating L dwarfs, which
can approach $v\sin i$ $\approx$ 60 km/s (Basri et al.\
2000), do not have strong emission, in contrast to hotter
rapidly-rotating stars which are typically the most active.

When they are young, however, brown dwarfs are apparently
not as extreme in their rotation velocities.  Joergens \&
Guenther (2001) have shown that substellar objects in Cha
I rotate as rapidly as their T-Tauri counterparts.
Basri (2002) has
proposed that the absence of magnetic activity, and
hence magnetic braking, allows brown dwarfs to retain
their angular momentum as they contract.  However, this
appealing hypothesis is complicated by the detection
of disks around young brown dwarf candidates (Muench et
al.\ 2001; Testi et al.\ 2002) and the presence of magnetic
flares even in the L dwarfs (Hall 2001).  Investigations
of rotation velocities in young cluster brown dwarfs are
currently underway to clarify this picture (G.\ Basri, 
priv.\ comm.). 
Theoretical studies of disk braking in these cool, low-mass
systems are desperately needed.

Photometric variability studies have also begun to measure
rotation periods in ultracool dwarfs, but results have been
somewhat confusing.  Work by Bailer-Jones \& Mundt (1999,
2001), Mart{\'{\i}}n et al.\ (2001), Clarke et al.\ (2002),
and Gelino et al.\ (2002) have found photometric variations
in late-M and L dwarfs on order 0.01--0.1 mag with aperiodic
or periodic (timescales of a few hours) modulations.
However, periods have been difficult to characterize in
general, as some objects show multiple and equally significant
periods (Mart{\'{\i}}n et al.\ 2001), while others exhibit
evolution in the photometric periods
(Bailer-Jones \& Mundt 2001; Gelino et al.\ 2002).

The difficulty in obtaining robust rotational periods
for these objects is likely related to the presence
of condensate clouds in their atmospheres (Ackerman \&
Marley 2001), which may produce rotationally modulated
features that evolve over a similar timescale (Gelino et
al.\ 2002).  Indeed, the transition from L dwarfs to T
dwarfs (Burgasser et al.\ 2002a; Geballe et al.\ 2002)
may be driven by the sudden disruption of cloud layers
by convective cells (Burgasser et al.\ 2002b), obscuring
any rotational modulation.  Various studies of photometric
variability are currently underway, however, because of the
feasibility of such projects on small (1-4m) telescopes,
in contrast to the high sensitivity required to measure
$v\sin i$ in these faint dwarfs.

\medskip


\noindent 
{\it Keivan G. Stassun}
\\ {\it University of Wisconsin---Madison}

\smallskip \noindent 
{\it Donald Terndrup} \\ 
{\it Ohio State University}

\end{document}